\documentclass{ieeeaccess}
\usepackage{cite}
\usepackage{amsmath,amssymb,amsfonts}
\usepackage{algorithmic}
\usepackage{graphicx}
\usepackage{textcomp}

\def\BibTeX{{\rm B\kern-.05em{\sc i\kern-.025em b}\kern-.08em
    T\kern-.1667em\lower.7ex\hbox{E}\kern-.125emX}}
\begin{document}
\history{Date of publication xxxx 00, 0000, date of current version xxxx 00, 0000.}
\doi{10.1109/ACCESS.2017.DOI}

\title{Bi-objective Optimization for Energy Aware Internet of Things Service Composition}
\author{\uppercase{Osama Alsaryrah}\authorrefmark{1}, 
\uppercase{Ibrahim Mashal\authorrefmark{2}, and Tein-Yaw Chung}.\authorrefmark{1},\IEEEmembership{Member, IEEE}}
\address[1]{Department of Computer Science and Engineering, Yuan Ze University, Chung-Li 32003,Taiwan}
\address[2]{Computer Science Department, Aqaba University of Technology, Aqaba 77110, Jordan }
\tfootnote{This work was supported by the Ministry of Science and Technology of Republic of China, Taiwan, under contract number MOST 106-2221-E-155-014.}

\markboth
{Alsaryrah \headeretal: Bi-objective Optimization for Energy Aware Internet of Things Service Composition}
{Alsaryrah \headeretal: Bi-objective Optimization for Energy Aware Internet of Things Service Composition}

\corresp{Corresponding author: Tein-Yaw Chung (csdchung@saturn.yzu.edu.tw).}

\begin{abstract}
In recent years, service-oriented based Internet of Things (IoT) has received massive attention from research and industry. Integrating and composing smart objects functionalities or their services is required to create and promote more complex IoT applications with advanced features. When many smart objects are deployed, how do we select the most appropriate set of smart objects to compose a service by considering both energy and Quality of Service (QoS) is an essential and challenging task. In this work, we reduced the problem of finding an optimal balance between QoS level and the consumed energy of the IoT service composition to a Bi-objective Shortest Path Optimization (BSPO) problem and used an exact algorithm named pulse to solve the problem. The BSPO has two objectives, minimizing the QoS including execution time, network latency, and service price and minimize the energy consumption of the composite service. Experimental evaluations show that the proposed approach has short execution time in various complex service profiles. Meanwhile, it can obtain good performance in energy consumption and thus network lifetime while maintaining a reasonable QoS level.
\end{abstract}


\titlepgskip=-15pt

\maketitle

\section{Introduction}
\label{sec:introduction}
\PARstart{I}{nternet} of things (IoT) is a new technology paradigm that allow smart objects to be connected together to collaborate, cooperate, and communicate with each other to provide and support smart applications \cite{b1,b2}. Currently, there are more than 8 billion connected smart objects, and the number will continue to increase dramatically year after year \cite{b3}. Smart objects are heterogeneous in their functionalities, communication capabilities, and resources. In general, smart objects are resource constrained with very limited computation and storage capacities when it is a battery-powered device, e.g., wireless sensors and mobile phones.

With the advent and rapid development of service-defined everything, smart objects are represented as services corresponding to their co-hosted functions \cite{b4,b5,b6}. In other words, each IoT smart object provides its function through standard services that can be directly accessed. To this end, Service-Oriented Computing (SOC) \cite{b6} are seen as the key enabler for IoT. 

Integrating and composing smart objects functions or their services is required to create and promote more complex IoT applications with advanced features \cite{b4}. Composing those services is done by aggregating atomic services to provide new functions that none of the services could provide individually \cite{b7}. This integration must consider the quality of service (QoS) and energy efficiency of the composed objects. Take a large-scale complex IoT environment as an example. When a smart object has low energy, it should be replaced with another smart object, if any, that has more energy and can provide the same functions and a good QoS level. However, this is a challenging task since IoT QoS values are dynamic and can substantially vary during the lifetime of the application when network states change. Moreover, smart objects can join, leave, fail, or new services with better quality can appear at any time. Therefore, finding a good balance between the energy consumption of all objects and its QoS to prolong the network lifetime is not a simple task.

In the past, several studies have discussed service discovery and composition and many techniques have been developed  for Web services and Representational State Transfer (REST) \cite{b8}. However, those techniques only considered and discussed the functional and non-functional properties of services. As mentioned above, due to the nature of IoT and the smart objects, IoT services composition must consider not only QoS, but also power consumption and residual energy level \cite{b9}. 

There are a few studies on IoT service composition that address energy consumption \cite{b10}. However, none of them considered energy consumption as a separated objective; they considered both QoS and energy consumption as a single objective. Unlike previous studies, we deal with energy consumption of IoT services separately from other QoS attributes. In this way, both QoS level and energy consumption are treated equally with the same priority.

This study aims to model and develop a Bi-objective Shortest Path Optimization (BSPO) for IoT service composition and to find an optimal balance between QoS level and the consumed energy of the IoT service composition. The BSPO energy-aware IoT service composition has two objectives: minimize the QoS including execution time, network latency, and service price and minimize the energy consumption of the composite service. BSPO derives the optimal solution by finding a Pareto-optimal solution for QoS and energy-aware IoT service composition based on users or operators preferences.

The main contribution of this paper can be summarized as follows:
\begin{enumerate}
\item Investigate IoT service composition by considering the energy consumption along with QoS of selected services.
\item Formulate a novel optimization problem to maximize the QoS level and to minimize the energy consumption of composite service. 
\item Propose a BSPO model and use the pulse algorithm to solve the formulated problem. 
\item Simulate and evaluate the proposed service composition scheme, and compare its performance with other greedy algorithms.
\end{enumerate}

The rest of the paper is organized as follows. Section II    reviews the related work. Section III presents the service composition model. Section IV describes problem formulation. Section V depicts the optimization problem. Section VI explains the pulse algorithm. Section VII shows results and performance evaluation. Finally, Section VIII concludes the paper.
\section{Related work}
\label{sec:Related work}
\subsection{Service-oriented IoT}
IoT adopts service-oriented architecture (SOA) paradigm because it can provide cooperation between heterogeneous smart objects and is very flexibile for system integration. Accordingly, smart objects can be connected and composed via service composition approaches. Recently, many research works have abstracted IoT devices as a service to provide an efficient and unified way of accessing and operating IoT services \cite{b1,b11,b12}.

The Web of Things (WoT) \cite{b13} has been proposed to seamlessly integrate heterogeneous objects.  Through existing Web technologies such as Web services and RESTful interfaces, WoT enable objects to communicate and interact. Sun et al.\cite{b14} proposed a microservice IoT framework to provide a generic IoT architecture based on a module or multicomponent application instead of the monolithic applications. The framework abstracts smart objects and IoT application modules as services.

Cheng et al.\cite{b15} proposed a platform for event-driven service-oriented IoT coordination, where SOA is adopted to solve interoperability issues among large numbers of heterogeneous services and physical entities in IoT. Another service-oriented IoT architecture is presented in \cite{b16}, where the authors proposed a user-centric IoT-based Service-Oriented architecture to integrates services that utilize IoT resources in an urban computing environment. In\cite{b16}, user goals are represented as an explicit task definition that is coordination of activities. Activities consist of configurations of abstract services that can be instantiated by orchestrating available service instances, including services that can be actuated through the IoT devices or composed of more than one smart object.

Many studies also focused on wireless sensor networks (WSNs) to provide composite services by integrating its functions. Zhou et al.\cite{b4} proposed a three tiers service-oriented framework. The function of each sensor is abstracted as a service within a service class. Service classes are chained to fulfill the functional requirements and energy-efficiency. Another work adopted service-oriented WSN presented in \cite{b17}.

\subsection{Service composition and QoS-aware services composition }
Service composition techniques are designed for relatively complex services when the required functions cannot be satisfied by any single service. It combines more than one service to fulfill the request. Traditional service composition techniques compose services in a specific order to meet the required goals\cite{b9}. In the literature, many techniques have been proposed for service composition based on its function such as semantic-based matchmaking, Logic, Graph-Theory, Petri net, and AI-Planning based \cite{b18}. 

QoS-aware services composition is known to be an NP-hard problem. However, this problem has already been addressed by several methods. Ngoko et al.\cite{b19} proposed a mixed-integer linear programming (MILP) method to solve QoS service composition. Furthermore, they considered energy consumption as a QoS attribute. Yu et al.\cite{b20} formulated the problem as a Multi-dimensional Multi-choice Knapsack Problem (MMKP). Llinás and Nagi \cite{b21} proposed a graph-based model to solve the problem as a Multi-Constraint Shortest Path problem (MCSP). Wu and Zhu \cite{b22} used a directed acyclic graph (DAG) to model services composition as a path search problem. Several studies proposed Pareto optimality techniques for solving the QoS-based services selection problem \cite{b23,b24}. For example, Chen et al.\cite{b23} proposed a services composition algorithm using a partial selection approach. Based on dominance relation, this approach allows us to reduce the search space by pruning unpromising candidate services in QoS. However, they adopted service selection by local optimization, which only selects the best candidate service locally for each abstract service without considering the relation between tasks in the workflow. Unlike the work cited above, we considered the QoS for the entire workflow as an end to end service composition and used an exact method to solve the bi-objective optimal problem including QoS and energy consumption.

\subsection{Service composition in IoT context}
Many studies in QoS-aware service composition and selection problem are available in the literature. Bellido et al. \cite{b7} analyzed stateless compositions of RESTful services and its control-flow patterns. The researchers also presented a comparative evaluation of different QoS attributes. Dar et al.\cite{b25} addressed the problem of integrating IoT smart objects by adopting the concepts of centralized service composition (orchestration) and decentralized service composition (choreography). 

Simple Additive Weighting (SAW) technique has been used to rank candidate IoT service in QoS. Kouicem et al.\cite{b26} and Yachir et al.\cite{b27} calculated a single utility value by aggregating the QoS values, where the service composition problem is transformed to a single objective optimization problem that finds the services with the best utility value. Jin et al.\cite{b24} proposed a three phases service composition algorithm. The first phase pre-sorts services according to user's preferences. The second phase applies dominance-based filtering to eliminate sub-optimal solutions and the final phase sorts the rest of services to select the best service. Four QoS attributes are associated with each IoT. The candidate services are evaluated concerning user's requirements by aggregating each QoS rating (utility) function. However, studies above do not consider energy consumption.

In the literature, only a few studies considered both energy and QoS. For example, Khanouche et al. \cite{b10} proposed an IoT energy-centered and QoS-aware services selection services and composition. The proposed selection approach preselects the services offering the QoS level required for user's satisfaction using a lexicographic optimization strategy and QoS constraints relaxation technique. The concept of relative dominance of services is also proposed. However, the preselecting phase aggressively prunes services that don’t meet the local QoS requirements or affect the quality of end to end QoS level when considering execution time and network latency. 

Unlike the aforementioned studies, our work deals with the energy efficiency of services separately from other QoS attributes and takes into account both QoS level and energy consumption of IoT services. The proposed selection approach transformed the problem into a bi-objective optimization problem that aims to minimize the amount of energy consumed and maximize the QoS of a service. We solved the optimization problem using bi-objective shortest path algorithm with four online pruning techniques to reduce the complexity of the algorithm. In addition, to ensure a good balance among the candidate services in energy consumption, the services consuming less energy are selected for each service composition. Our model aims to provide high availability of services by minimizing the energy consumption, i.e., by maximizing the devices battery lifetime. 

To the best of our knowledge, there exists no previous study that considers an end to end service composition scheme with both QoS and the energy efficiency as the primary metrics for IoT resources.
\section{IOT SERVICE COMPOSITION MODEL}
\label{IOT SERVICE COMPOSITION MODEL}
IoT applications consist of a set of decomposed services. In our model, we define two types of services: an abstract service $\mathrm{(}AS\mathrm{)}$ represents function of a service provided by one or more IoT devices or software modules, while a candidate service ($cs$) represents a real Web service that may be invoked. All $cs\mathrm{\ }$are distributed across available resources.

The candidate services are characterized by two types of properties: functional and non-functional properties. Functional properties indicate the actions and the functions provided by a$\mathrm{\ }cs$, while non-functional properties are defined by the QoS attributes and the energy profile of the service running at a battery-powered object. Here, we considered three QoS attributes, namely, execution time, cost, and energy consumption profile.

From a user's perspective, when a user's service request arrives at a service composition broker, the service composition process is invoked. The process combines a series of atomic service components appropriately to form a composite service (path) that provides an optimal balance between QoS and consumed energy.

Fig.1 shows a model for composing IoT services. Suppose~an IoT service consists of n~tasks, $\{T_i,~1\le i\le n\}$, which are needed by a given IoT service requested by an end user. For each task $T_i$ a corresponding abstract service, ${AS}_i$, is used to represent its functional requirement. For each ${AS}_i,1\le i\le n,$~, there exists~$l_i$ candidate services,${\ \{CS}_{i,j},~1\le j\le l_i$, that can meet the functional requirements of~${AS}_i$~and thus can be selected for realizing this service component. Note that a service component may be either an IoT service or a software component. To meet the requirements of a specific service request, the service composition path is constructed as a path from a service entrance portal (${CS}_0$) to a service exit portal CSe by traversing only one candidate service of each service component. For example, ~${CS}_0$$\mathrm{\to}$${CS}_{1,3}$$\mathrm{\to}$${CS}_{2,5}$$\mathrm{\to}$$\mathrm{\cdots }$$\mathrm{\to}$${CS}_{m,j}$$\mathrm{\to}$${CS}_e$~is a composite service path that may provide useful service to a user. Therefore, the main focus for IoT services composition is to select an optimal service sequence from a pool of available smart objects and software components while satisfying various QoS requirements, in addition to the amount of consumed energy.
\Figure[t!]( topskip=0pt, botskip=0pt, midskip=0pt){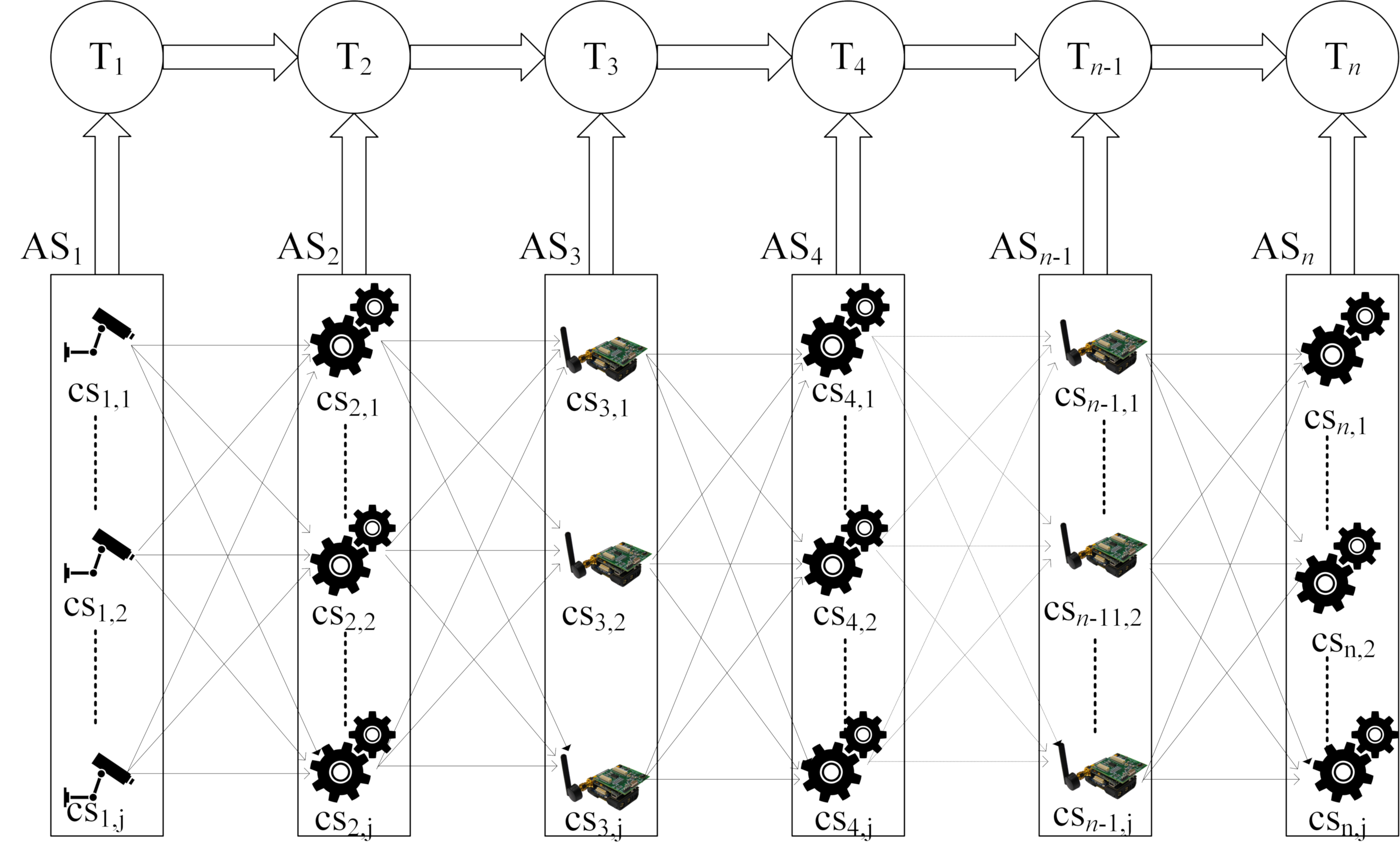}
{Service composition model\label{fig1}}
\section{Problem formulation}
\label{sec:Problem formulation}

Let  $T\mathrm{=}\mathrm{\{}T_{\mathrm{1}},T_{\mathrm{2}},T_j\mathrm{,\dots ..,}T_{\mathrm{n}}\}$ denote the set of tasks that cover the composite IoT service, where n is the total number of decomposed tasks and $T_j$ is the j${}^{th}$ (j = 1, 2, 3, n) sub task of T. Let${\ AS}_i\mathrm{=}\mathrm{\{}{cs}_{i\mathrm{1}},{cs}_{i\mathrm{2}},{cs}_{ij}\mathrm{,\ }{cs}_{im_{\mathrm{i}}}\}$ be the candidate services available for task$\ T_j$, where $m_{\mathrm{i}}$ represents the number of candidate services and ${cs}_{ij}$ is the j${}^{th}$ candidate service. Thus, the directed graph model shown in Fig.1 represents all possible compositions for an IoT service.

A distributed IoT service formed by interconnecting n different service components can be modeled as a directed graph~$G\mathrm{=(}N,A\mathrm{)}$~, where~$N\mathrm{=}\mathrm{\{}v_{\mathrm{1}}\mathrm{,\dots ,}v_i\mathrm{,\dots ,}v_n\}$ denotes a set of service components with n nodes and~$A\mathrm{=}\mathrm{\{}e_{ij}\mathrm{\ |}v_i\mathrm{,\ }v_j\mathrm{\in }\mathrm{\ }N\mathrm{\ }\mathrm{\}}$~is the set of edges (links). Each node $v_i$ is associated with a weight $c_i$ representing its functional capability of abstract service${\ AS}_i$, $1\le i\le n$. Each edge $e_{ij}\mathrm{\in }A$ are associated with two nonnegative weights~${QoSU}_{ij}$~and$\mathrm{\ }{EP}_{ij}$, where~${QoSU}_{ij}$~and$\mathrm{\ }{EP}_{ij}$ denote the utility value of QoS attributes and the consumed energy respectively when traversing$\mathrm{\ }e_{ij}$. Henceforth, without loss of generality, ${QoSU}_{ij}\ $refers to QoS attributes such as execution time, network latency and cost. The objective of service composition is to select one candidate service from each set ${AS}_j$ and generate an optimal IoT Composition Service Path (CSP) $x\mathrm{=}\mathrm{\{}x^{\mathrm{1}},x^{\mathrm{2}}\mathrm{,\dots ..,}x^i\mathrm{,\dots \dots ,}x^{\mathrm{n}}\}$  from the set of available compositions under multi-objective requirements, where $x^i$ denotes the candidate service selected for sub task$\ T_i$.

\subsection{Energy Profile Model (EP)}
Since energy consumption is a significant factor for devices hosting the candidate service, it is considered necessary that each candidate service provides the composer its energy consumption variable so that the composer can select the most energy efficient one.

We defined the energy profile (EP) of ${cs}_{ij}$ by two variables, the residual energy level ${RE\mathrm{(}cs}_{ij}\mathrm{)}$ of the device hosting ${cs}_{ij}$and the consumed energy ${CE\mathrm{(}cs}_{ij}\mathrm{)}$ which represents the consumed energy when running ${cs}_{ij}$. ${RE\mathrm{(}cs}_{ij}\mathrm{)}$ is estimated as follows.

\begin{equation}
{RE\mathrm{(}cs}_{ij}\mathrm{)}  ={CDE\mathrm{(}cs}_{ij}\mathrm{)-\ }{E_{th}\mathrm{(}cs}_{ij}\mathrm{)}\label{eq1}
\end{equation}

where~${CDE\mathrm{(}cs}_{ij}\mathrm{)\ }$ represent current energy level of the battery-powered device hosting ${cs}_{ij}$ and ${{\mathrm{\ }E}_{th}\mathrm{(}cs}_{ij}\mathrm{)}$ the energy threshold value under which the device cannot support ${cs}_{ij}$ anymore.

Since our model is based on service-oriented computing, the consumed energy of running ${cs}_{ij}$, ${CE\mathrm{(}cs}_{ij}\mathrm{)}$,~is calculated as in Eq. \eqref{eq2}. Here, we assumed that the energy consumption of ${cs}_{ij}$ is constant since the service runs on the same platform, uses the same resources, and receives and sends the same amount of data.:

\begin{equation} \label{eq2} 
{CE\mathrm{(}cs}_{ij}\mathrm{)=}ECR\left({cs}_{ij}\right)\mathrm{*}T\mathrm{(}{cs}_{ij}\mathrm{)} 
\end{equation} 
where~$ECR\left({cs}_{ij}\right)$ represents the energy consumption rate, and~$T\mathrm{(}{cs}_{ij}\mathrm{)}$~represent the execution time of ${cs}_{ij}$.
The energy profile of ${cs}_{ij}$~is calculated as shown in Eq. \eqref{eq3} by taking the ratio between the energy consumed by invoking ${cs}_{ij}$ and its residual energy.

\begin{equation} \label{eq3} 
{EP\mathrm{(}cs}_{ij}\mathrm{)}= {\mathrm{\ }CE\mathrm{(}cs}_{ij}\mathrm{)/\ }{RE\mathrm{(}cs}_{ij}\mathrm{)}
\end{equation}

Thus, the smaller is ${EP\mathrm{(}cs}_{ij}\mathrm{)}$ the better is the$\mathrm{\ }{cs}_{ij}$ as an candidate for $T_i$. 

Finally, the EP of a service composition $x$ can be calculated as shown in Eq. \eqref{eq4}, the energy consumed by composition path $x^i$ is:

\begin{equation}\label{eq4}
EP\left(x\right)\mathrm{=}\sum^n_{i=1}{EP\left(\left.x^i\right|x^i\in S^h\right)} 
\end{equation}

where $S^h$ represents the set of battery-powered devices. In our model, we consider the differences in the underline infrastructure of the running services. In general, IoT software services are not executed on battery powered devices in contrast to sensing and actuating services. To differentiate between these two types of required services we refer to the set of battery-powered devices by${\mathrm{\ }S}^h$.

\subsection{QoS Criteria}
In our study, we consider a set of quantitative non-functional properties of IoT services which can be used to describe the quality criteria of a web service. For the sake of simplicity, in this paper, we consider only negative attributes as our non-functional properties. These values of negative attributes need to be minimized. We included two attributes: Service Execution Time (T), which is collected from records of previous execution monitoring, and Service Cost (C), which is directly collected from service providers.

\subsubsection {Service Execution Time (T):}

Service execution time represents the average time expected for executing a candidate service. The service provider updates it continuously because the load of the device hosting the service changed dynamically. Clients expect their jobs to be completed in a minimal time when they submit requests to the service composer. Let ${\mathrm{\ }L\mathrm{(}cs}_{ij}\mathrm{)}$~be the latency of transmitting data from ${cs}_{ij}$ and $ET\left({cs}_{ij}\right)\ $be the execution time of service request for ${cs}_{ij}$. Thus, the service execution time T(${cs}_{ij}$) can be computed as in Eq. \eqref{eq5}.

\begin{equation}\label{eq5}
{T\mathrm{(}cs}_{ij}\mathrm{)} ={\mathrm{\ }L\mathrm{(}cs}_{ij}\mathrm{)+}ET\mathrm{(}{cs}_{ij}\mathrm{)}.
\end{equation}
Therefore, the service execution time for composition path $x$ can be given as follows.
\begin{equation} \label{eq6}
T\left(x\right)\mathrm{=\ }\sum^n_{i\mathrm{=1\ }}{T\mathrm{(}x^i)\mathrm{}} 
\end{equation}

\subsubsection {Service Cost (C):}
When a user submits his/her request to a service composer, the composer manages and finds the fastest composition path for him/her. Meanwhile, the user is also expected to pay the fairest price for running his/her tasks. Therefore, the Service Cost is considered a valuable QoS property. In this model, we set $C\mathrm{(}{cs}_{ij}\mathrm{)}$ as the cost of executing${\mathrm{\ }cs}_{ij}$. The cost is usually fixed but may be changed according to the service provider's business policy. The execution cost is registered by the service provider.

Therefore, as given in Eq. \eqref{eq7}, the service cost for composition path $x$ is:
\begin{equation} \label{eq7}
C\left(x\right)\mathrm{=\ }\sum^n_{i\mathrm{=1}}{C\left(x^i\right)} 
\end{equation}

\subsubsection {Utility Function}
Using utility function is a helpful mechanism for evaluating the aggregated quality of a given composite service. In this research, we calculate the utility value of a service composition by aggregating normalized QoS attributes values. All QoS values are mapped to a single real value between 0 and 1 by comparing the QoS value with the minimum and maximum available QoS value. This enables uniform evaluation of the QoS value.

We adopted a Multiple Attribute Decision-Making approach, namely, the Simple Additive Weighting (SAW) technique\cite{b28} for the mapping process. For a composite service path \textit{x} the aggregated QoS values are compared with the minimum and maximum possible aggregated values. The minimum (or maximum) possible aggregated values can be easily estimated by aggregating the minimum (or maximum) value of each service class. 

The utility function of QoS is computed as in Eq. \eqref{eq8}.

\begin{equation} \label{eq8}
QoSU\left(x\right)\mathrm{=}w_t\mathrm{*}\acute{T}\left(x\right)\mathrm{+}w_c\mathrm{*\ }\acute{C}\left(x\right)
\end{equation}

where$\mathrm{\ }w_t$and $w_c$ represent the weighting factors of execution time and cost respectively. The sum of weights are equal to one, i.e., $\mathrm{\ }w_t\mathrm{+}$ $w_c$ $\mathrm{=1}$. $\acute{T}\left(x\right)$ is the normalized service execution time for \textit{x} that is calculated using~Eq. \eqref{eq6}.~$\acute{C}\left(x\right)$~is the normalized service cost for \textit{x}  that is calculated using~Eq. \eqref{eq7}.

With the above QoS utility and power profile formulas, we can formulate the optimization problem of the service composition as a BSPO problem. We described the bi-objective shortest path optimization problem in~the next section.

\section{Bi-objective optimization problem}
\label{Bi-objective optimization problem}
Multi-objective service composition selects one candidate service from each set ${AS}_j$ and generates an optimal IoT CSP from the set of available compositions under multi-objective and constraints. In this study, we formulated the energy aware IoT service composition as BSPO for finding an optimal IoT CSP ~from the start node~${cs}_s\mathrm{\in }N$~to the end node~${cs}_e\mathrm{\in }N$~that minimizes two different (often conflicting) objective functions. The bi-objective shortest path problem can be formally defined as in Eq. \eqref{eq9}:

\begin{equation}\label{eq9}
\begin{split}
{\mathrm{min} CSP\mathrm{(}x\mathrm{)}\ }\mathrm{=(}QoSU\mathrm{(}x\mathrm{),}EP\mathrm{(}x\mathrm{))}\\
\begin{aligned}
\text{s.t.,\qquad} \qquad
\\  & x\mathrm{\in }X \qquad \qquad
\end{aligned}
\end{split}
\end{equation}

where~$\boldsymbol{x}$~ represents a candidate service path from the service entrance portal from~${cs}_s\ $to the service exit portal${\mathrm{\ }cs}_e$. $QoSU(x)$ represents the aggregated value of the QoS values along all edges of a path $x$. $EP\left(x\right)$ represents the aggregated value of the $EP$ over all edges in $x$.~$X$~is the set of all paths from~${cs}_s$~to${\mathrm{\ }cs}_e$. The objective of~Eq. \eqref{eq9}~ is to minimize the QoS utility value and the energy profile of CS$P\mathrm{(}x\mathrm{)}$. Since the existence of a path that simultaneously minimizes both objectives in~Eq. \eqref{eq9}~cannot be guaranteed, we seek for a set of paths with an acceptable tradeoff between the two objectives. 

IoT service composition is a NP-complete problem with $\prod^n_{j\mathrm{=1}}{M_j}$  possible CSP for task T, where \textit{n} represents the number of tasks and $M_i$ represents the number of candidate services for task $T_i$. To solve the above problem, in this paper, we use the pulse algorithm detailed in the next section.

\section{The pulse algorithm: overview}
\label{The pulse algorithm: overview}
We used the pulse algorithm \cite{b29} to solve the bi-objective optimization problem. The pulse algorithm optimizes a bi-objective function CSP(\textit{x}) that is composed of a quality of service function $QoSU$ and energy profile function \textit{EP}. A path $x$ optimizes CSP or is Pareto-optimal if there is no other path $x'$ that has lower \textit{QoSU} and lower \textit{EP} than$\ x$. The goal of the pulse algorithm is then to find a series of such Pareto-optimal paths that together form an efficient set${\mathrm{\ \ }X}_E$ by recursively examining the entire search space of the graph. 

The efficiency of the algorithm is achieved by aggressively pruning partial paths. When a pulse reaches to a newly added node, it will check if adding the node to the existing partial path satisfies one of the adopted pruning conditions or not. The partial path will be eliminated if it meets one of the following conditions:

\begin{enumerate}
\item It includes cycles.
\item It exceeds either one or both upper bounds obtained at the initialization phase, represented by a nadir point, before reaching the end node.
\item It is dominated by any other path in the current efficient set before reaching the end node.
\item It is dominated by any objective value stored in the label of the newly added node. A node is labeled by accumulated objective values when it is traversed by a feasible path. Thus, if the partial path is dominated by the existing label of the newly added node, it will not be part of a Pareto optimal path.
\end{enumerate}

Suppose that we have a given network with start node~$v_s$~~and end node${\mathrm{\ }v}_e$.The pulse algorithm sends a pulse from $v_s$~~to ${\mathrm{\ }v}_e$. This pulse travels through the entire network while storing the~partial path $p$~(an ordered sequence of visited nodes) and its cumulative objective functions,~$QoSU\left(p\right)$ and$\mathrm{\ }EP\mathrm{(}p\mathrm{)}$. Every pulse that reaches the end node${\mathrm{\ }v}_e$ is a feasible solution that might be efficient. Once a pulse reaches the end node, it recursively backtracks to continue its propagation through the rest of the nodes in the search for more efficient paths from~$v_s\mathrm{\ }$to${\mathrm{\ }v}_e$. If the pulse is let free, this recursive algorithm identifies all possible paths, and guarantees that an efficient set is always found. 

However, the pulse algorithm does not continue exploring any partial path that will not produce an efficient solution by using a look-ahead mechanism~that prunes~aggressively vast regions of the solution space.~For the~initialization~procedure, the algorithm starts running a mono-objective shortest path algorithm to get the upper bound for each objective. The pulse algorithm is shown in Algorithm 1.

 The pulse algorithm follows a depth-first search truncated by several pruning strategies to control the pulse propagation and prunes pulses without cutting off any efficient solution. The algorithm defines four pruning strategies namely, pruning by cycles, nadir point, efficient set, and~label. The pulse~recursive function is shown in~Algorithm 2~. It takes four input parameters, current node${\mathrm{\ }v}_i$, the cumulative QoS utility value$\ QoSU\mathrm{(}p\mathrm{)}$,~the cumulative energy profile$\ EP\mathrm{(}p\mathrm{)}$,~and the partial path~\textit{p}. The pruning strategies applied to the pulse are shown in Lines 1--4; if the pulse is not pruned, line 5 stores the current~$QoSU\mathrm{(}p\mathrm{)}$~and~$EP\mathrm{(}p\mathrm{)}$~and line 6 adds the node~$v_i$~to the partial path. In lines 7--11, the pulse propagates over all nodes~$v_j$~$\mathrm{\in }$~${\mathit{\Gamma}}^{\mathrm{+}}\mathrm{(}v_i\mathrm{)}$, where~${\mathit{\Gamma}}^{\mathrm{+}}\mathrm{(}v_i\mathrm{)}$ is the set of outgoing neighbors of~$v_i$, and adds current~$QoS\ $~utility to the cumulative one and current~$EP$~to the cumulative $EP$.
 
Whenever the pulse function is invoked at end node${\mathrm{\ }v}_e$, a partial path~$P$~becomes a complete solution~$x$~and we update the online efficient set$\widehat{{\mathrm{\ }X}_E}$. Note that the information about~$X_E$~has a global scope and is not an attribute of the traveling pulse within the recursion.~Algorithm 3~presents the pulse function when it is invoked on end node${\mathrm{\ }v}_e$. Since a new solution has been found, the algorithm verifies if the new solution is efficient and updates the online efficient set accordingly.
\begin{table}
\setlength{\tabcolsep}{3pt}
\begin{tabular}{p{3in}}
\hline
Algorithm 1: Pulse algorithm \\ \hline 
Input: $G$ directed graph;$\mathrm{\ }{\mathrm{\ }v}_s$: start node; ${\mathrm{\ }v}_e$ end node\newline Output: ${\mathrm{\ }X}_E$: true efficient set\newline 1: $p\mathrm{\leftarrow }\mathrm{\{}\mathrm{\ }\mathrm{\}}$\newline 2: $QoSU\mathrm{(}p\mathrm{)}\mathrm{\leftarrow }\mathrm{0}$\newline 3: $EP\mathrm{(}p\mathrm{)}\mathrm{\leftarrow }\mathrm{0}$\newline 4: $initialization\left(G\right)$\newline 5:$\mathrm{\ }pulse\mathrm{(}{\mathrm{\ }v}_s,QoSU\left(p\right),EP\left(p\right),p\mathrm{)}$\newline 6: return ${\mathrm{\ }X}_E$ \\ \hline 
\end{tabular}
\end{table}
\begin{table}
\setlength{\tabcolsep}{3pt}
 \begin{tabular}{p{3.2in}}\hline
 Algorithm 2: Pulse function$\mathrm{:}\ \newline
 pulse\left({\mathrm{\ }v}_s,QoSU\left(p\right),EP\left(p\right),p\right)$  \\ \hline 
 
Input: ${\mathrm{\ }v}_i,current\mathrm{\ }node\mathrm{;}\mathrm{\ }QoSU\left(p\right)\mathrm{,\ }\newline
cumalative\mathrm{\ }QoS\mathrm{\ }utility\mathrm{;}$ $EP\left(p\right)\mathrm{,\ }\newline
cumulative\mathrm{\ }power\mathrm{;}\mathrm{\ }p\mathrm{,\ } current\mathrm{\ }path.$\newline
Output: void\newline
1: if isAsyclic(${\mathrm{\ }v}_i,p$) then \newline
2: \hspace{2pt} if checkNadirPoint(${\mathrm{\ }v}_i,QoSU\left(p\right),EP\left(p\right)$) then    \newline
3: \hspace{8pt} ifcheckEfficientSet(${\mathrm{\ }v}_i,QoSU\left(p\right),EP\left(p\right)$) then\newline
4:\hspace{16pt} ifcheckLabels(${\mathrm{\ }v}_i,QoSU\left(p\right),EP\left(p\right)$) then\newline
5:\hspace{22pt} store ($QoSU\left(p\right),EP\left(p\right)$)\newline
6:\hspace{25pt}$p\mathrm{\leftarrow }\mathrm{\ }\grave{p}\mathrm{\cup }\mathrm{\{}{\mathrm{\ }v}_i\}$   \newline  
7:\hspace{22pt} for ${\mathrm{\ }v}_j\mathrm{\in }{\mathit{\Gamma}}^{\mathrm{+}}\mathrm{(}v_i\mathrm{)}$  do\newline  
8:\hspace{33pt} $qos\left(\grave{p}\right)\mathrm{\leftarrow }QoSU\left(p\right)\mathrm{+}QoSU\mathrm{(}{cs}_{ij}\mathrm{)}$\newline  9:\hspace{33pt} $EP\left(\grave{p}\right)\mathrm{\leftarrow }EP\left(p\right)\mathrm{+}EP\mathrm{(}{cs}_{ij}\mathrm{)}$\newline 
10:\hspace{30pt}  $pulse\mathrm{(}{\mathrm{\ }v}_j,QoSU\left(\grave{p}\right),EP\left(\grave{p}\right),\grave{p}\mathrm{)}$\newline 
11:\hspace{18pt} end for\newline 
12:\hspace{12pt} end if   \newline 
13:\hspace{8pt} end if\newline 
14:\hspace{6pt} end if  \newline 
15: \hspace{1pt} end if\newline 
16: return void
\\ \hline
 \end{tabular}
 \end{table}
 \begin{table}
\setlength{\tabcolsep}{3pt}
\begin{tabular}{p{3.2in}} \hline 
Algorithm 3: Pulse function for the end node$\ :\mathrm{\ \ }\ pulse\left({\mathrm{\ }v}_e,QoSU\left(p\right),EP\left(p\right),p\right)$        \\ \hline 
Input: ${\mathrm{\ }v}_e,current\mathrm{\ }node\mathrm{;}\mathrm{\ }QoSU\left(p\right)\mathrm{,\ } \newline cumalative\mathrm{\ }QoS\mathrm{\ }utility\mathrm{;}$ $EP\left(p\right)\mathrm{,\ } \newline cumulative\mathrm{\ }power\mathrm{;}\mathrm{;}\mathrm{\ }p\mathrm{,\ }current\mathrm{\ }path.$\newline 
Output: void\newline 
1: if CheckEfficientSet(${\mathrm{\ }v}_e,QoSU\left(p\right),EP\left(p\right)$) then\newline 
2:\hspace{8pt}$\mathrm{\ }p\mathrm{\leftarrow }\mathrm{\ }p\mathrm{\cup }\mathrm{\{}{\mathrm{\ }v}_e\}$   \newline 
3:\hspace{8pt} $\boldsymbol{X}\mathrm{\leftarrow }\mathrm{mapPathToSolution(}p\mathrm{)}$\newline
4: \hspace{8pt} UpdateEfficentSet(X)\newline 
5:end if\newline 
6:return void \\ \hline 
\end{tabular} 
 \end{table}
\subsection{Pruning techniques}
\subsubsection{Pruning by Cycles:}
Because all weights on the arcs are nonnegative, any efficient solution cannot contain cycles. To avoid cycles in a path, every time we invoke the pulse function at${\mathrm{\ }v}_i$, the algorithm checks whether a node has been visited or not. If~${\mathrm{\ }v}_i$  has already lain on the partial path,~it~is pruned from P.

\subsubsection{Pruning by Nadir point:}
Based on the idea of the nadir point which seen as the anti-ideal point in the objective space, the algorithm aims to prune as early as possible any pulse exceeding either smallest values (best path) of both objectives solutions. To do so, we calculate the minimum \textit{QoSU(x)} (regardless of EP) and the minimum energy profile \textit{EP(x) }(regardless of QoS) from entrance node to the end node.

Assume that we have an optimal solutions $x^{\mathrm{*}}_{QoSU}$ and ~$x^{\mathrm{*}}_{EP}$ where~$x^{\mathrm{*}}_{QoSU}$~and~$x^{\mathrm{*}}_{EP}$~represent Energy profile and QoS utility objectives of the mono-objective shortest path problem respectively. The images for the optimal solutions in the objective space are~$Z\left(x^{\mathrm{*}}_{QoSU}\right)\mathrm{=(}\overline{EP},\underline{QoSU}\mathrm{)}$ and$\mathrm{\ }Z\left(x^{\mathrm{*}}_{EP}\right)\mathrm{=(}\underline{EP},\overline{QoSU}\mathrm{)}$. The nadir point, denoted by $Z^N\mathrm{=(}\overline{EP},\overline{QoSU}\mathrm{)}$ ,  which represent a vector composed of the worst objective values in the objective space. In other words, it represents an upper bound for each objective in the objective space.
 
 After applying the mono objective shortest path problem for each objective, a set of an alternative optimal solution for each objective can be found. ~$\overline{EP}$~and $\mathrm{\ }\overline{QoSU}\mathrm{\ }$represent the smallest values among all alternative solutions for both objectives~$x^{\mathrm{*}}_{QoSU}$~and~ $x^{\mathrm{*}}_{EP}$~,~respectively. As shown in Fig.2  $Z\left(x^{\mathrm{*}}_{QoSU}\right)$ , $Z\left(x^{\mathrm{*}}_{EP}\right)$, and $Z^N$ shows the minimizer victors in the objective space and  Z* represents the ideal point.
 
 Based on this, for any solution~\textit{x}~with~ $QoSU\left(x\right)\mathrm{>\ }\overline{QoSU}\mathrm{\ }$ or$\mathrm{EP}\left(x\right)\mathrm{>}\overline{EP}$ , its image~z(\textit{x}) is dominated and~\textit{x}~is not efficient. Where any point falls in the dark region in fig 2. Is eather dominated by $Z\left(x^{\mathrm{*}}_{QoSU}\right)$ or $Z\left(x^{\mathrm{*}}_{EP}\right)$.based on this any pulse exceeding either~$\overline{EP}$~and $\mathrm{\ }\overline{QoSU}$ will be pruned. 
 
 \Figure[t!]( topskip=0pt, botskip=0pt, midskip=0pt){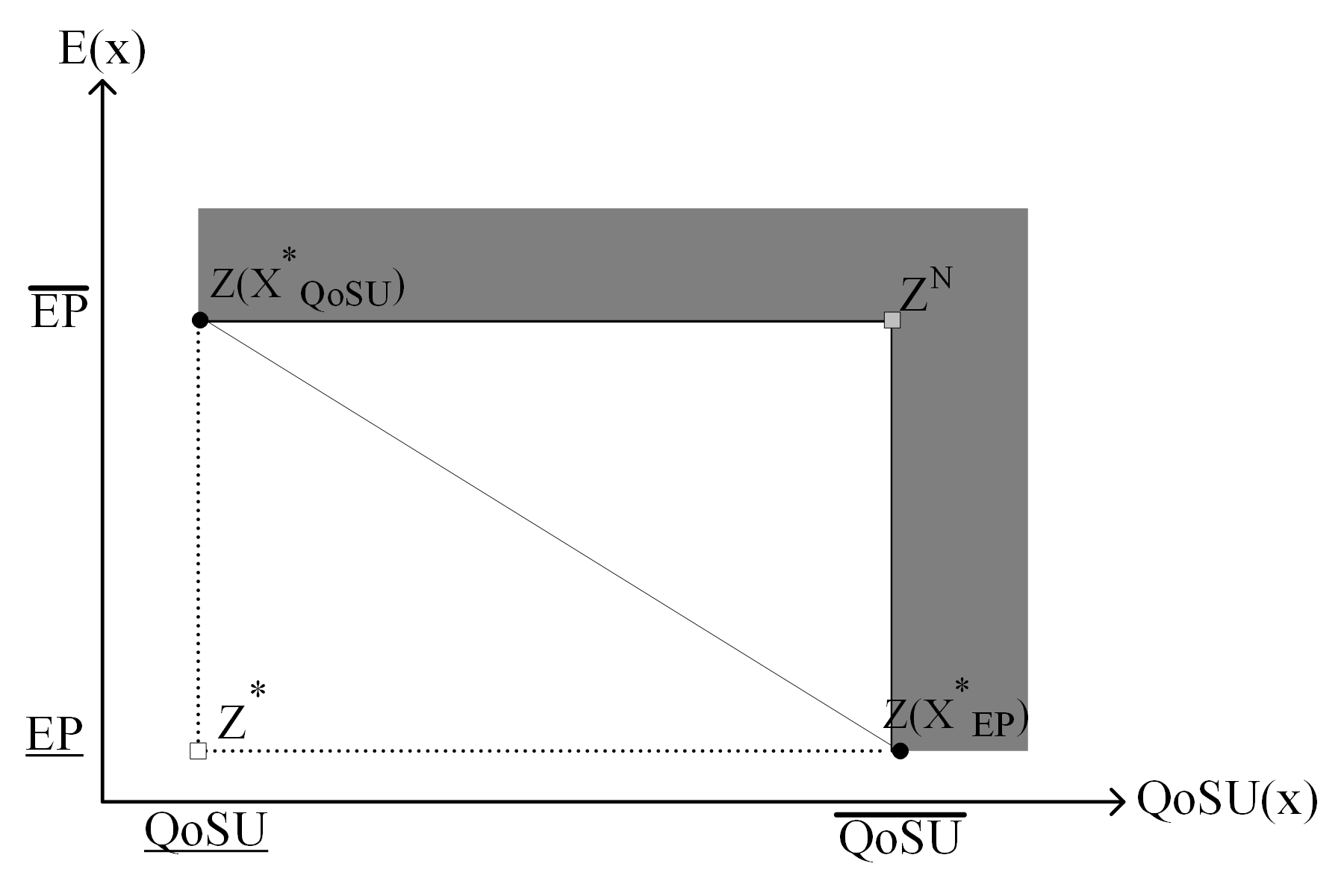}
{Nadir point and lower and upper bounds\label{fig2}}
 
 \subsubsection{Pruning by Efficient set:}
 Consider the online efficient set$\widehat{{\mathrm{\ }X}_E}\ $at a given intermediate stage of the algorithm. Using the lower bound found in the initiation phase of the algorithm namely,~$\underline{QoSU}\mathrm{(}v_i\mathrm{)}$ for QoS utility and~$EP\left(v_i\right)$ for energy profile, we can determine whether a partial path will become an efficient solution or not.  Given a partial path~\textit{p}~to node$\ v_i$, if there is a solution$\ x\mathrm{\in }\widehat{{\mathrm{\ }X}_E}$ such that $QoSU\left(p\right)\mathrm{+\ }\underline{QoSU}\mathrm{(}v_i\mathrm{)}\mathrm{\ge }QoSU\mathrm{(}x\mathrm{)}$  and \\ $\mathrm{\ }EP\left(p\right)\mathrm{+\ }\underline{EP}\mathrm{(}v_i\mathrm{)}\mathrm{\ge }EP\mathrm{(}x\mathrm{)}$, we can safely prune partial path~\textit{p},~because even if it spends both the minimum cost and the minimum time to reach the end node, it will still be dominated by~path \textit{x}.
 
 \subsubsection{Pruning by label:}
 For each node~$v_i$ a fixed number of labels saves a tuple of \textit{QoSU} and \textit{EP} values. The labels at node~$v_i$~are denoted by~$\mathrm{L}\left({\mathrm{v}}_i\right)\mathrm{=}\left\{\left({QoSU}_{il},{EP}_{il}\right)\right|l\mathrm{=1,\dots ,Q}\mathrm{\}}$ where ~${QoSU}_{il}$~and~${EP}_{il}$are the cumulative QoS utility and energy profile for a partial path to~$v_i$ respectively and~Q~denotes the number of labels at${\ v}_i$. For an incoming pulse, the algorithm checks if the incoming partial path~\textit{p}~is dominated or not; that is, if any label dominates~$CSP\mathrm{(}p\mathrm{),}$~the pulse is discarded by label pruning.
\section{Performance Study}
\label{Performance Study}
In this section, we presented the setup of our simulation. Then we analyzed the performance of the exact bi-objective algorithm for IoT service composition that optimizes QoS utility, i.e. network latency, response time and service cost, and consumed energy.  

To show how IoT services composition can be instantiated and invoked while keeping an optimal balance between QoS level and the consumed energy of the composed service, we conducted an extensive simulation under various scenarios to evaluate the performance of the proposed IoT service composition. In these scenario we assumed that we have a smart environment consist of thousands of heterogeneous objects such as mobile devices, wireless sensors, and smart home and smart building devices such as smart power outlets, shatters, elevators, access controls, air conditioning, surveillance cameras, etc. We also assumed that these devices are heterogeneous in communication protocols. For example, wireless sensors can be based on ZigBee, 6lowpan or Bluetooth [2]. To provide interoperability between these heterogeneous objects the functions of these objects and software components are abstracted as services to become accessible through SOP, COAP  or REST protocols [2]. Finally, we assumed that all candidate services are registered in IoT orchestration system and categorized based on its services classes.

\subsection{Simulation Environment and Methodology}
We implemented our simulation using java under 64-bit Windows seven operating system, running on Intel Core i5-2500, 3.3 GHz, and 8 GB RAM. Each scenario generated a different number of services classes AS and candidate services \textit{cs}. Each candidate service has two QoS attributes: execution time including network latency, and service cost.

Due to the absence of datasets in QoS and energy profile values of IoT services, we chose to evaluate the proposed algorithm by using synthetically generated data. For instance, each service is produced with a random QoS value according to the values reported in the literature [10]. Based on the model presented in [29]\textbf{,} we specified the energy profile of IoT services. The values of each quality parameter are generated according to a normal distribution. The importance or weight of each QoS attribute in the utility function is set to be fixed 1/2. The service execution time and network latency are generated assuming a uniform distribution over the interval [20, 1200] and [20, 800]. The cost of services is generated according to a uniform distribution over the interval [10, 20].

In order to overcome the problem of QoS values fluctuation during service runtime in dynamic IoT environments, QoS values are randomly changed after every service iteration by multiplying every QoS value with a random number in the interval [0.9, 1.1]. 

In order to study the energy consumption in battery operated objects, we referred to the energy model presented in [29]. We assumed that every device has an initial amount of charge and maximum battery charge, $Cinitial$ and$\ cmax$, where the values of $Cinitial$ is chosen randomly in the interval [0.7 $Cmax$,1.0 $Cmax$] and the maximum battery  charge $Cmax$ of an object is set as 1500 mA$\mathrm{\cdot }$h. Furthermore, any object has energy lower than $CThreshold$ becomes unable to provide its services and will not be considered in the composition process. In this study, we set $CThreshold$ to be 30\% of$\ cmax$. After every run of the selected candidate service, a specific amount of power is consumed and this amount is subtracted from current battery level of the device hosting the service. The amount of consumed power is chosen randomly in the interval [100 mA.s, 10000 mA.s] after every service invocation.

In our study, a service composer is used to find an optimal balance between QoS and consumed energy and prolong the network life time. In the following sections we will refer to the proposed algorithm by BOSC (Bi-Objective Service Composition). In order to show the added value of the proposed selection approach in a large-scale IoT services environment, we compared our results with two variants of our algorithm: QoSC, only the QoS is taken into account in the selection process, and EPC, only the power profile is taken into account in the selection process. The results presented here are derived based on the average of 100 simulations.

\subsection{Simulation Results}
To evaluate the performance of the proposed model and algorithm we considered the following metrics: (1) Selection time which represents the computational time of the selection algorithm; (2) Energy consumption of the composite service which is equal to the total energy consumed by its components; (3) Composition lifetime which is the number of compositions that can be executed before the first candidate service failure. A service is considered failed when its autonomy is no longer sufficient to be invoked; (4) Optimality which is the ratio between the QoS value of the composite service obtained by BOSC and the optimal QoS value of the composite service, obtained by that of QoSC and EPC.
 \subsubsection{Selection Time versus Number of Services}
 To validate the scalability of BOSC, we tested the execution time of the selection algorithm under various numbers of tasks involved in the composition process and various numbers of available candidate services for each task. 

In the first experiment, we set the number of tasks involved in the composition to be 10, 15 and 20 tasks. We also set the number of candidate services for each task to be between 100 and 1000. Fig. 3 compares the average execution time (in millisecond or \textit{ms}) of the composition algorithm with various numbers of tasks and candidate services for each task. As shown in Fig 3, the average execution time increases as the number of tasks of the composite service increases. As shown in the figure, the average execution time is short and suitable for a large scale IoT environment. For example, the selection time does not exceed 10 \textit{ms} when running 10 tasks each with 1000 candidate services. When increasing the number of tasks to 20 each with 1000 candidate services, the average execution time increases slightly to reach less than 70 \textit{ms}. However, this increase is still reasonable and acceptable.
 \Figure[t!]( topskip=0pt, botskip=0pt, midskip=0pt){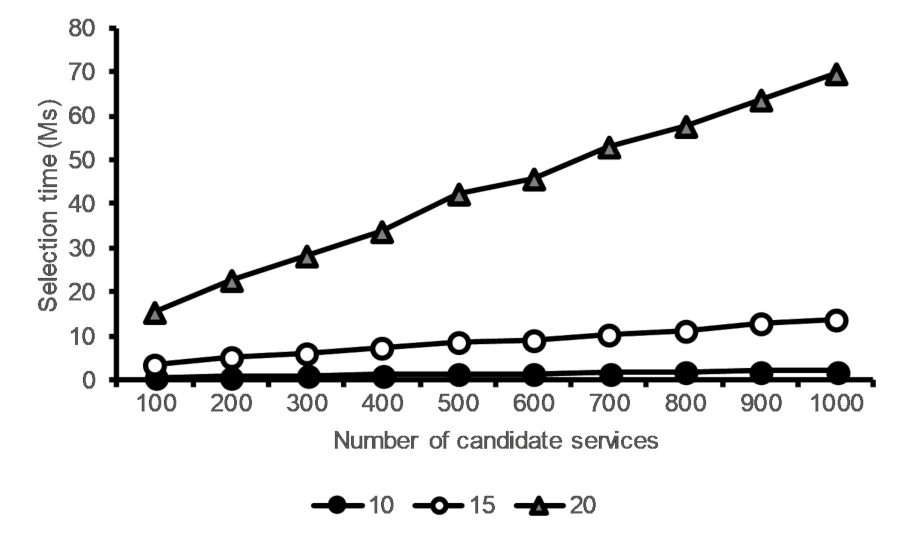}
{Selection time versus number of candidate services (10, 15, 20 tasks\label{fig3}}

\subsubsection{Energy Consumption Versus Number of Services}
In the second simulation, we compared the performance of the three algorithms in consumed energy (in mA.s). In the simulation, we considered a composition path consisting of 10 tasks where each task has 100 to 1000 candidate services. As shown in Fig. 4, the amount of consumed energy in BOSC and EPC gets close to each other. Note that in EPC the candidate services with the lowest energy profile are always selected. 

From~Fig. 4,~we~can~see~that when candidate services are between 100 and 1000, the amount of consumed energy by BOSC is about 35\% more than that by EPC in average. Certainly, EPC provides the best composite service in energy consumption. An interesting observation is that the amount of consumed power decreases when the number of the available services increases. This can be explained by the fact that when increasing the number of candidate services the probability of selecting more services with less power consumption increases, and hence, the energy consumed decreases. Another advantage of BOSC is that it saves more power than that by QoSC. BOSC consumed energy 70\% less than that by QoSC.
 \Figure[t!]( topskip=0pt, botskip=0pt, midskip=0pt){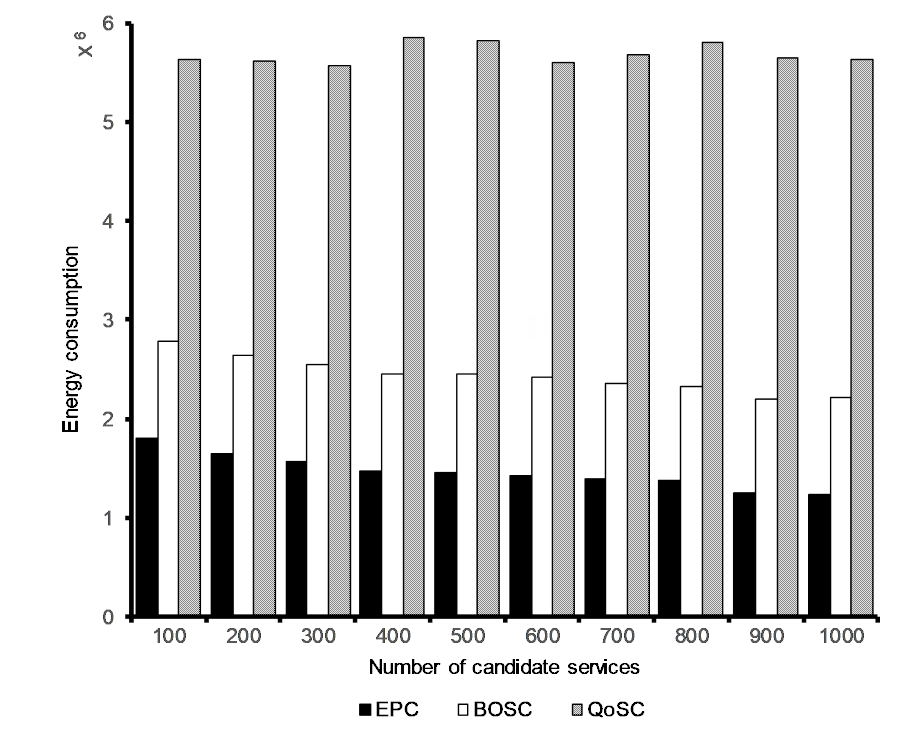}
{Energy consumption versus number of candidate services.\label{fig4}}

In the third experiment, we intended to show the consumed energy under various numbers of tasks between 10 and 50 tasks when the number of candidate services for each task are set to be 100. Fig. 5 shows that the energy consumption of the composite path increases when the number of service classes increases. In fact, increasing the size of composite path will increase the number of selected services which causes more power consumption.

\Figure[t!]( topskip=0pt, botskip=0pt, midskip=0pt){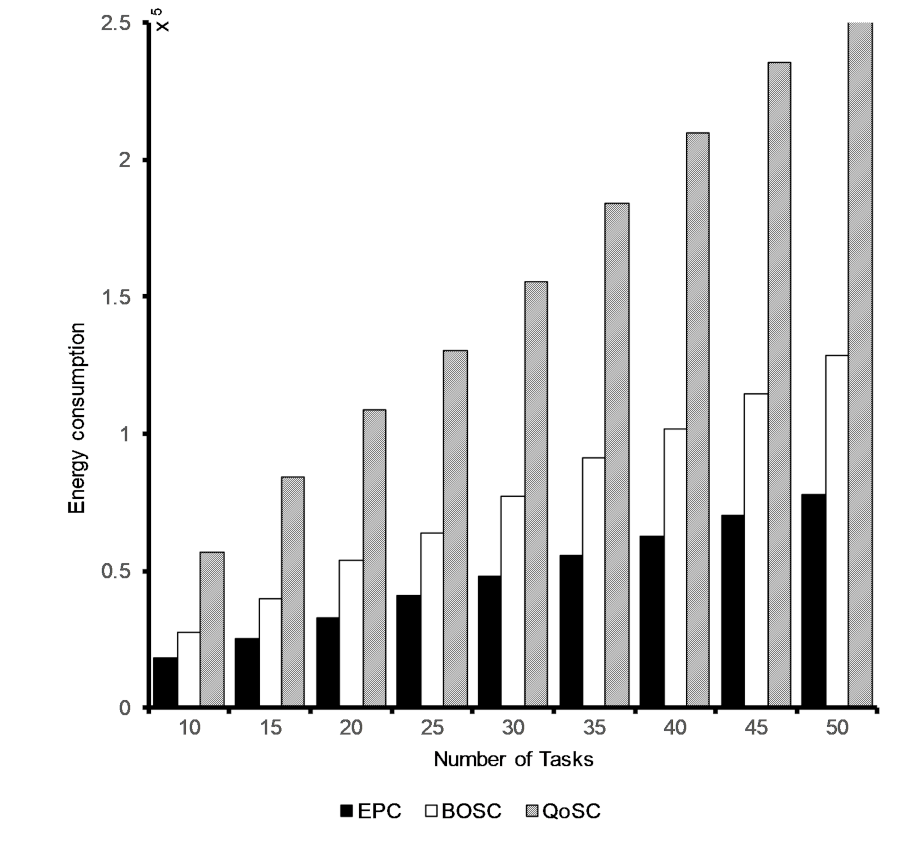}
{Energy consumption versus number of tasks)\label{fig5}}
\subsubsection{Composition Lifetime Versus Number of Concrete Services}

Studying the performance of the proposed algorithm in service composition life time is the aim of the fourth experiment. We evaluated and compared the composition lifetime by BOSC with that by EPC and QoSC. In the simulation, we considered a composition path consisting of 10 tasks and each task has 100 to 1000 candidate service. As shown in Fig. 6, the composition life time with BOSC is slightly less than that by EPC when only the lowest energy profile is selected. On the other hand, EPC guarantees the lowest energy consumption while reducing the \textit{QoSU} of the composite path. Indeed, BOSC can achieve a good balance between the amount of consumed energy and \textit{QoSU} of the composed path. Thus, BOSC ensures a long composition life time, which provides a high availability of candidate services. 
\Figure[t!]( topskip=0pt, botskip=0pt, midskip=0pt){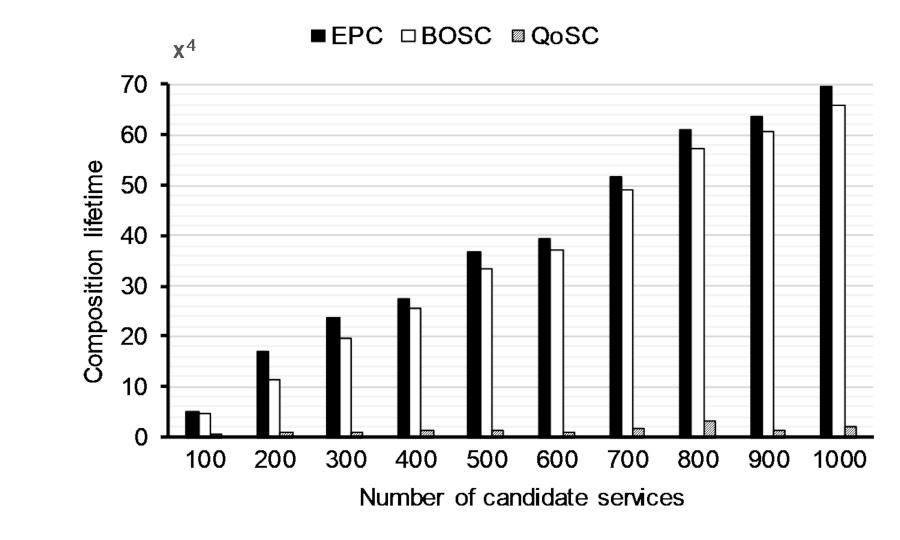}
{Composition lifetime versus number of candidate services.\label{fig6}}

\subsubsection{Optimality of the Solution}
Studying the performance of BOSC in optimality of the obtained \textit{QoSU} is the purpose of this experiment.  In the simulation we considered a composed path consisting of 10 tasks and each task has 100 to 1000 candidate services. As shown in Fig. 7, the optimality of the proposed method guaranteed a QoS level about 80\% close to that acquired by QoSC. It is worthy of noting that BOSC does not apply any constraint on QoS attributes in the simulation. The results of BOSC can be further improved if we apply some constraints on QoS attributes, such as execution time and cost thresholds. This is because it helps to reduce the solution choices with lower \textit{QoSU} and hence better QoS can be achieved. As we can observe from Fig. 7, the QoS optimality level of EPC dramatically decreases, while BOSC can provide a solution close to optimal solutions.
\Figure[t!]( topskip=0pt, botskip=0pt, midskip=0pt){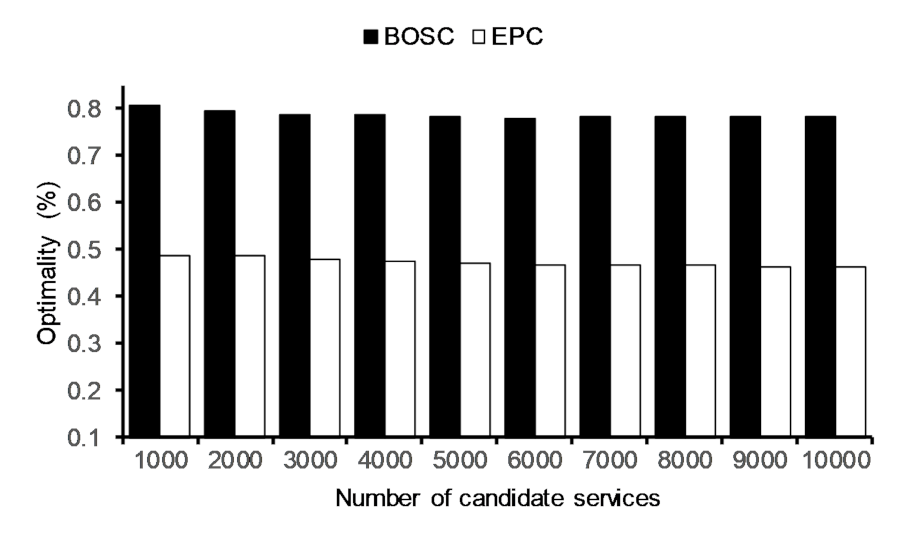}
{Optimality QoS versus number of candidate services.\label{fig7}}

\section{CONCLUSION}
\label{CONCLUSION}
Service-oriented IoT has received~considerable attention over the past~few years. A~crucial~factor~for~the~success~of IoT and its applications is to create more complex IoT applications with advanced features by composing smart objects functions and services.

In this paper, a bi-objective shortest path optimization model is presented to model IoT service composition where energy consumption and QoS are considered. The pulse algorithm with four embedded pruning techniques, namely, pruning by cycle, nadir point, efficient set and label, is developed to solve efficiently the presented problem. Results show that our proposed IoT service composition scheme overcomes and surpasses other schemes that only consider QoS or power consumption individually. Experiments also show that the proposed scheme works reasonably fast in selecting suitable smart objects; the average execution time needs less than 70 ms, which makes the proposed model scalable for large-scale IoT environments. The amount of consumed energy by BOSC is about 35\% more than that consumed by EPC on average. The composition lifetime with BOSC is 90\% more than that by the QoS only scheme. Also, the acquired optimality level of the BOSC guaranteed a QoS level about 80\% close to that obtained by QoSC. Therefore, the proposed solution provides an optimal balance between QoS level and consumed energy in IoT service composition.

\begin{IEEEbiography}[{\includegraphics[width=1in,height=1.25in,clip,keepaspectratio]{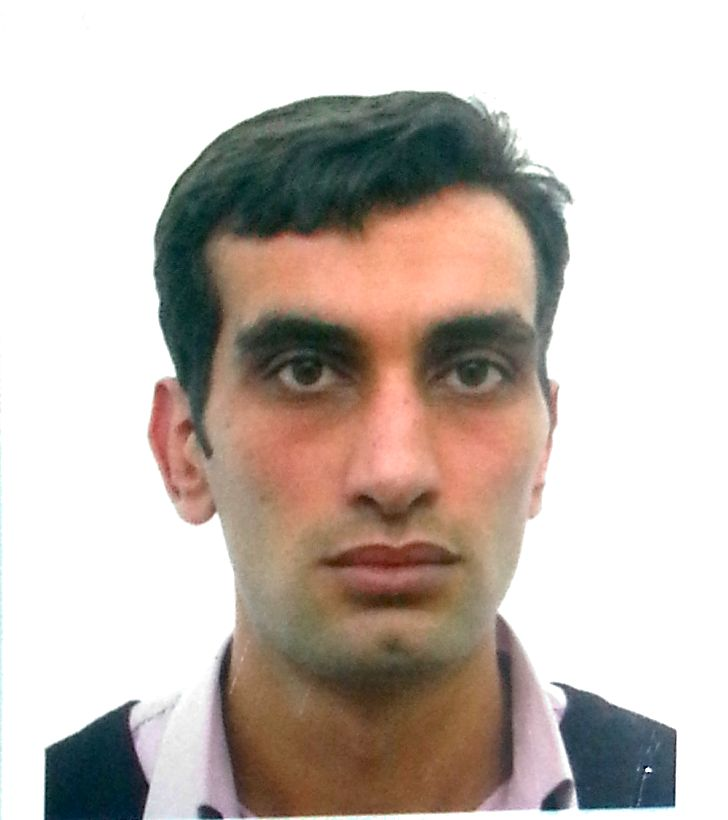}}]{Osama Alsaryrah}  is a Ph.D. candidate in the Department of Computer Science and Engineering, Yuan Ze University, Taiwan. In 2007 he received his M.S. in computer science from the University of Jordan, Jordan. His research interests include future communications systems, Internet of Things and Smart Applications, Web services, Sensor networks and Distributed computing
\end{IEEEbiography}

\begin{IEEEbiography}[{\includegraphics[width=1in,height=1.25in,clip,keepaspectratio]{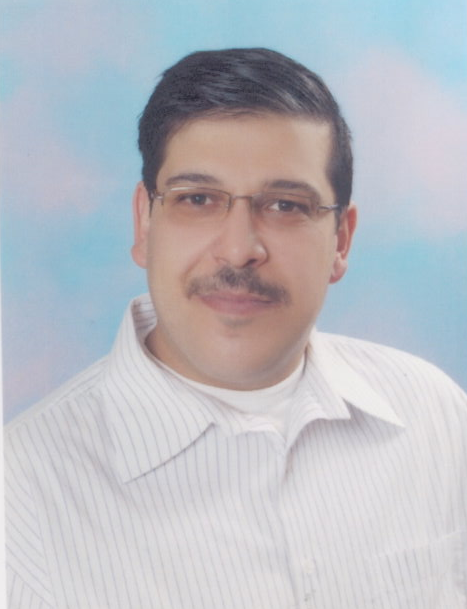}}]{Ibrahim Mashal}  is a faculty member in the Computer Science department at Aqaba University of Technology. He earned his PhD. degree in the field of Computer Science, specializing in Internet of Things, from Yuan Ze University, Taiwan (R.O.C.) in 2016. He received his M.S. in Computer Science from the University of Jordan, Jordan. His research interests include Future communications systems, Internet of Things and Smart Applications, Wireless mobile network, and Sensor networks.  
\end{IEEEbiography}

\begin{IEEEbiography}[{\includegraphics[width=1in,height=1.25in,clip,keepaspectratio]{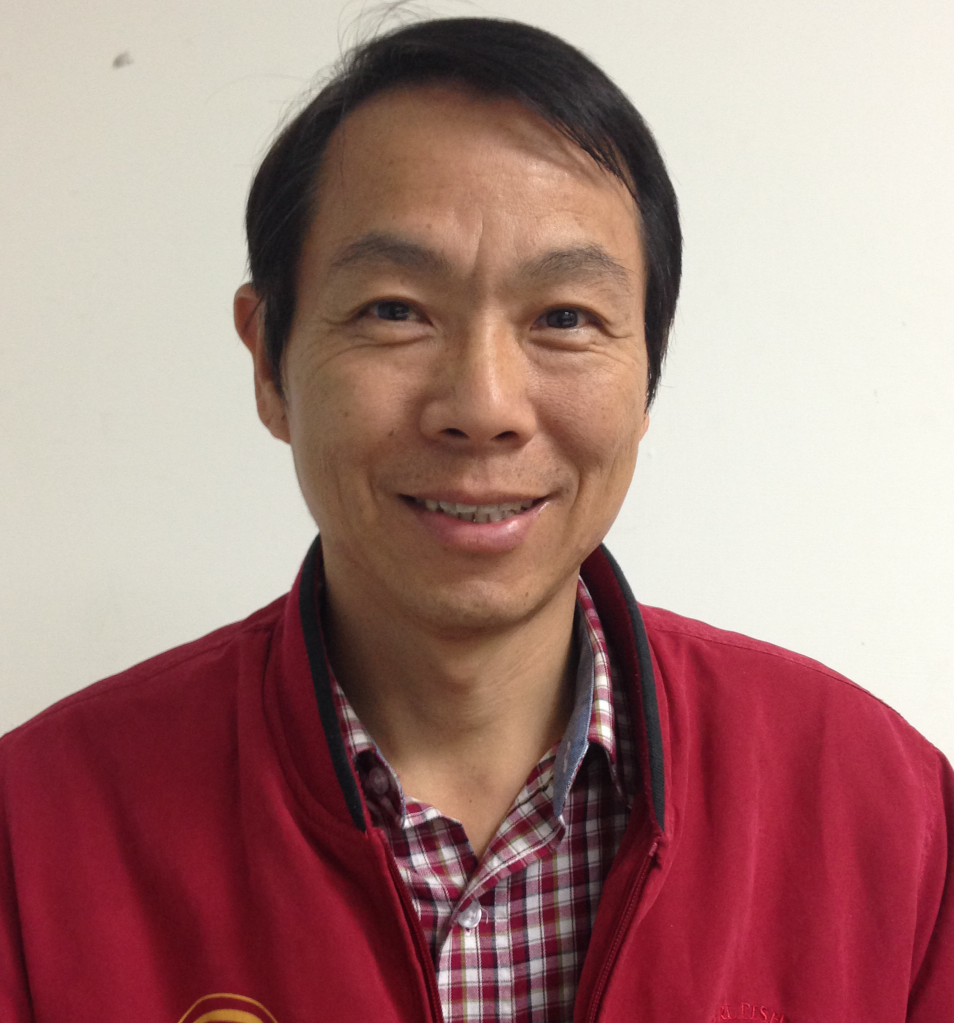}}]{Tein-Yaw Chung} (M'87) received the M.S. and Ph.D degrees in Electrical and Computer Engineering from North Carolina State University, Raleigh, NC, USA, in 1986 and 1990, respectively. From Feb.1990 to Feb 1992, he was with Network Service Division, IBM, RTP, NC, USA. Since May 1992, he has been with Yuan Ze University, Chung-Li, Taiwan and is a full Professor now. His current research interests include overlay network, Multimedia Communication and mobile networking. He is a member of IEEE.
\end{IEEEbiography}

\EOD

\end{document}